\documentclass[prl,twocolumn,superscriptaddress]{revtex4-1}

\usepackage{soul}
\usepackage{amssymb}
\usepackage{amsmath}
\usepackage{bbold}
\usepackage{braket}
\usepackage{xcolor}

\usepackage{hyperref}

\hypersetup{
    colorlinks,
    linkcolor={red!80!black},
    citecolor={green!70!black},
    urlcolor={blue!80!black}
}

\usepackage{graphicx}

\newcommand{\bigO}{\mathcal{O}}

\begin{document}


\title{Self-induced glassy phase in multimodal cavity quantum electrodynamics}

\author{V. Erba}
\thanks{ORCID Vittorio Erba: 0000-0001-8017-928X}
\email[\\Corresponding author: ]{vittorio.erba@posteo.net}
\affiliation{Dipartimento di Fisica dell' Universit\`a degli Studi di Milano, via Celoria 16, 20100 Milano, Italy}
\affiliation{Istituto Nazionale di Fisica Nucleare, sezione di Milano, via Celoria 16, 20100 Milano, Italy}

\author{M. Pastore}
\thanks{ORCID Mauro Pastore: 0000-0002-2771-7437}
\affiliation{Dipartimento di Fisica dell' Universit\`a degli Studi di Milano, via Celoria 16, 20100 Milano, Italy}
\affiliation{Istituto Nazionale di Fisica Nucleare, sezione di Milano, via Celoria 16, 20100 Milano, Italy}
\affiliation{Université Paris-Saclay, CNRS, LPTMS, 91405, Orsay, France}

\author{P. Rotondo}
\thanks{ORCID Pietro Rotondo: 0000-0001-6745-6166}
\affiliation{Dipartimento di Fisica dell' Universit\`a degli Studi di Milano, via Celoria 16, 20100 Milano, Italy}
\affiliation{Istituto Nazionale di Fisica Nucleare, sezione di Milano, via Celoria 16, 20100 Milano, Italy}

\begin{abstract}

We provide strong evidence that the effective spin-spin interaction in a multimodal confocal optical cavity gives rise to a self-induced glassy phase, which emerges exclusively from the peculiar euclidean correlations and is not related to the presence of disorder as in standard spin glasses.
As recently shown, this spin-spin effective interaction is both non-local and non-translational invariant, and randomness in the atoms positions produces a spin glass phase. Here we consider the simplest feasible disorder-free setting where atoms form a one-dimensional regular chain and we study the thermodynamics of the resulting effective Ising model. We present extensive results showing that the system has a low-temperature glassy phase. Notably, for rational values of the only free adimensional parameter $\alpha=p/q$ of the interaction, the number of metastable states at low temperature grows exponentially with $q$ and the problem of finding the ground state rapidly becomes computationally intractable, suggesting that the system develops high energy barriers and ergodicity breaking occurs.
\end{abstract}

\maketitle

Cavity quantum electrodynamics (CQED) provides --together with trapped ions \cite{PhysRevLett.74.4091,RevModPhys.75.281}, circuit \cite{PhysRevA.69.062320} and waveguide QED \cite{PhysRevLett.111.090502}-- one of the state-of-the-art controllable platform for quantum simulation. In a first series of experiments  \cite{PhysRevLett.107.140402,Baumann:Nat:10} it was shown that Bose-Einstein condensates in single-mode optical cavities undergo an abrupt change from a dark to a superradiant state, which is the non-equilibrium counterpart of the phase transition predicted by Hepp and Lieb (HL) \cite{HeppL73,HeppL:PRA:73,PhysRevE.67.066203,PhysRevLett.90.044101,Nagy:PRL:10} within their exact statistical physics analysis of the Dicke model.

More recently, a setting where the optical cavity sustains many degenerate electromagnetic modes was explored \cite{Lev2009NatPhys}. Here, the physics arising from the interaction of the many modes with the atoms in random positions is much richer than the one observed in the single-mode case \cite{Goldbart:PRL:2011,Strack:PRL:2011}: first, frustration (generated by the random positions of the atoms in the cavity) may lead to a glassy behavior; second, the form of the effective spin-spin couplings is reminiscent of the interaction arising from Hebbian learning in the so-called Hopfield model \cite{Hopfield2554}, that describes the simplest way to obtain an associative memory, i.e. a physical system with the ability to store and retrieve memory patterns from corrupted information. 


This discovery led the authors of Refs. \cite{PR2015PRL, PR2015PRB} to investigate the equilibrium statistical physics of the full quantum disordered Dicke model with $M$ degenerate electromagnetic modes, by generalizing the original work by HL. This analysis allowed to establish that, in the strong coupling limit, the energy landscape shares the same structure of minima of the Hopfield model, i.e. a ferromagnetic landscape with $M$ degenerate ground states corresponding to the stored memory patterns. Whether this picture survives to the interaction with an environment has been the subject of subsequent work \cite{PR2020PRL, PR2020PRR, PR2018JPA, carollo2020}.  

Notably, the feasibility of a quantum optical-based associative memory has been investigated in a series of remarkable papers by Lev, Keeling and coworkers \cite{Lev2019PRL,Lev2019PRA,Lev2018PRX}. Here the authors proved that the effective spin-spin interaction can be sign-changing with a tunable range and identified a practical protocol to implement associative memories in CQED \cite{Lev2020arxiv}. 

More importantly, they were able to obtain the specific form of the interaction for confocal cavities \cite{Lev2020arxiv}, showing explicitly its non-local and non-translational invariant nature (in fact, it depends on the scalar product of the positions of two atoms). Understanding the physics arising by this peculiar spin-spin interaction is a challenge by itself, since taking into account euclidean correlations is commonly a hard task in statistical physics. 



Our goal in this manuscript is to shed light on the effective spin-spin interaction arising in multimodal optical cavities by revealing an exotic spin glass phase that appears at low temperature \emph{without} any explicit quenched disorder in the Hamiltonian.
We consider the simplest feasible setting: atoms lie on a regular one-dimensional chain, collinear with the main axis of the optical cavity. In this case, the energy depends on a single adimensional parameter $\alpha$. For rational $\alpha =p/q$, the interaction is periodic and we show that the free-energy landscape is a $q$-dimensional manifold. Surprisingly, as $q$ grows, we observe an exponential proliferation of metastable states which is typical of disordered systems with a complex free-energy landscape. Our analysis shows that in multimodal CQED the presence of disorder in the atomic positions is not fundamental to achieve such a rough landscape. More formally, this physical system provides an explicit realization of self-induced quenched disorder as introduced in Refs. \cite{Bernasconi1987, Parisi1994JPA1, BouMez1994}.

\emph{The model---} We consider an Ising model with energy $E(\pmb \sigma) = \frac{1}{2N} \sum_{i,j=0}^{N-1} J_{ij} \sigma_i \sigma_j$ where the $\sigma_i$'s are binary $\pm 1$ variables and the couplings depend on the positions $\mathbf{r}_i$ of the atoms in the cavity as:
\begin{equation}
J_{i,j} = \cos\left(2\pi\frac{\mathbf{r}_i\cdot \mathbf{r}_j}{w_0^2}\right)\,,
\label{Eq.couplings}
\end{equation}
where the length scale $w_0$ is proportional to the width of the Gaussian TEM$_{00}$ mode. This form of the effective spin-spin interaction has been derived in detail in Refs. \cite{Lev2018PRX, Lev2019PRA, Lev2019PRL}. In \cite{Lev2020arxiv} the authors investigated, mostly with numerical methods, the properties of the resulting energy landscape, focusing on the case where atoms were in random positions. 
Here we consider a disorder-free scenario where atoms form a one-dimensional lattice, i.e. the $m$-th atom is in position $\mathbf{r}_m = a (m-L) \hat{\mathbf{n}}$ where $a$ is the lattice spacing, $L = N/2$ and $\hat{\mathbf{n}}$ is a unit vector orthogonal to the cavity mirrors and origin at the center of the cavity. 
Notice that this choice implies that the $L$-th atom lies at the origin of the chain; since the interaction is non-translationally invariant, other conventions may qualitatively change the physics of the system (see SM for more details \cite{SM}).
It is worth noticing that in this setting the couplings in Eq. \eqref{Eq.couplings} depend on a single adimensional parameter $\alpha = (a/w_0)^2$.

\begin{figure}[t]
\includegraphics{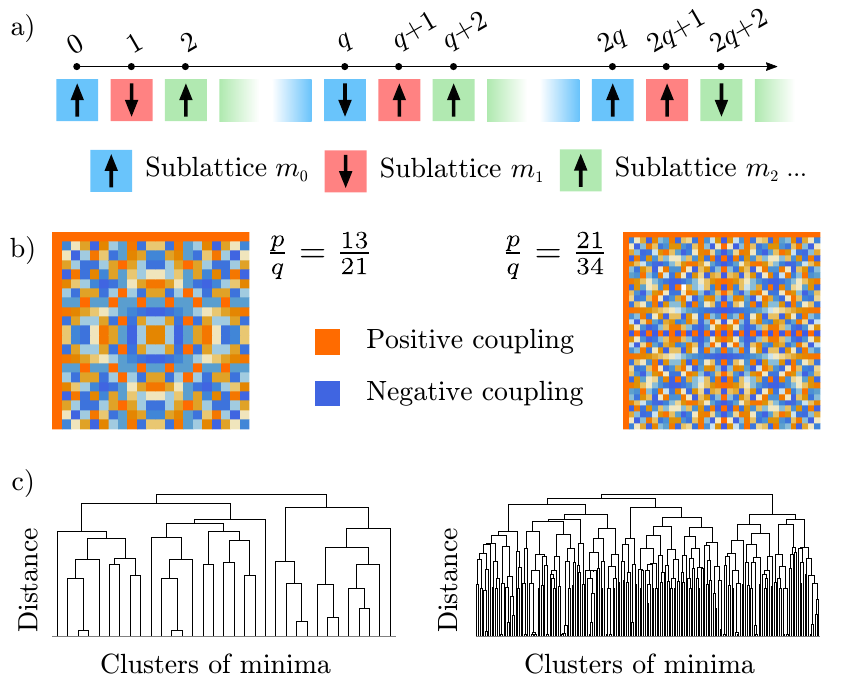}
\caption{\textbf{Collective variables, reduced interaction matrix and proliferation of metastable states increasing $\mathbf{q}$.} (a) Identification of the collective variables to study the partition function analytically. Since the interaction is periodic, it is natural to group spins lying on the sublattices $\Lambda_r = \{i q + r, i = 0,1,\dots, 2\ell -1\}$. In this way we can define $q$ mesoscopic magnetizations $m_r$, with $r=0,1,\dots,q-1$ and reduce the partition function to an integral over these $q$ variables. (b) Reduced $q \times q$ interaction matrix for two different values of $\alpha = p/q$. Removing the first row and column, the resulting $(q-1) \times (q-1)$ matrix displays a reflection symmetry around the anti-diagonal, that allows to further reduce the dimensionality of the free-energy manifold from $q$ to $\lfloor q/2 +1 \rfloor$. (c) Schematic representation by hierarchical clustering of the energy landscape at the two different values of $\alpha= 13/21, 21/34$. We obtain this representation by displaying the local minima of the energy in Eq. \eqref{eq. energy} as the leaves of a dendrogram. Distance is increasing along the vertical axis and pairs of distinct clusters at distance $d$ (measured from their center) are iteratively joint together in a new node at height $d$. As $q$ increases, we observe a proliferation of local minima. 
} 
\label{fig:1}
\end{figure}

\emph{Variational free-energy of the system---} Our first goal is to investigate the thermodynamics of the Ising model with couplings given by Eq. \eqref{Eq.couplings} by looking at its free-energy $f = -1/(N \beta) \log Z$, where $\beta = 1/T$ is the inverse of the temperature $T$ and $Z = \sum_{\pmb \sigma} e^{-\beta E(\pmb \sigma)}$ is the partition function. 
    We stress that the thermodynamic temperature $T$ that we introduce here is not the real temperature of the experimental implementation of the model.
    In fact, in \cite{Lev2020arxiv} it is proven that the relaxation dynamics of the system is a steepest descent dynamics in the free-energy landscape at null thermodynamic temperature $T$.
    For this reason we will focus on the case $T=0$ in the main text, and we will discuss the high-temperature phase in the SM \cite{SM}.

Remarkably, analytical progress can be made for rational $\alpha = p/q$, where $p$ and $q$ are coprime positive integers. For irrational $\alpha$, it is reasonable to expect that studying the rational approximation obtained by truncating the continued fraction expansion of $\alpha$ may deliver insights on the physics of the model, as it occurs, for instance, in Ising systems with long-range repulsion \cite{PhysRevLett.49.249,PR2016PRL} or in the Frenkel-Kontorova model \cite{AUBRY1983381}.

The strategy to compute the partition function of the model at rational $\alpha = p/q$ is the following: since the interaction is periodic with period $q$, it is convenient to introduce $q$ collective variables $m_r$ ($r = 0,1, \dots, q-1$) which represent the magnetizations on the sublattices $\Lambda_r = \{i q + r, i = 0,1,\dots, 2\ell -1\}$ (where for simplicity, $L = \ell q$, see also SM), i.e. $m_r = 1/(2\ell) \sum_{i=0}^{2 \ell -1} \sigma_{i q + r}$ (see also Fig. \ref{fig:1}). Notice that $|m_r| \leq 1$ $\forall r = 0,1,\dots, q-1$. This choice of grouping the microscopic variables allows to write the energetic contribution to the partition function as \cite{SM}:
\begin{equation}
\label{eq. energy}
E (\{m_r\}) = \frac{\ell}{q} \sum_{r,s=0}^{q-1} \cos\left(\frac{2\pi p}{q} r s \right) m_r m_s\,.
\end{equation}
The reduced $q \times q$ interaction matrix is depicted in Fig. \ref{fig:1} for two different values of $\alpha =p/q$. The entropic contribution can be obtained as well by counting the degeneracy of each magnetisation $m_r$, i.e. how many microscopic spin configurations contribute to a given value of the magnetisation. In the thermodynamic limit we can easily estimate this entropy as:
\begin{equation} 
\begin{split}
S(\{m_r\}) = 2\ell \left[ q \log 2 -\sum_{\sigma=\pm 1} \sum_{s=0}^{q-1}  \frac{1+\sigma m_s}{2}\log(1+\sigma m_s)   \right].
\end{split}
\end{equation}
 
\begin{figure*}[t]
\includegraphics{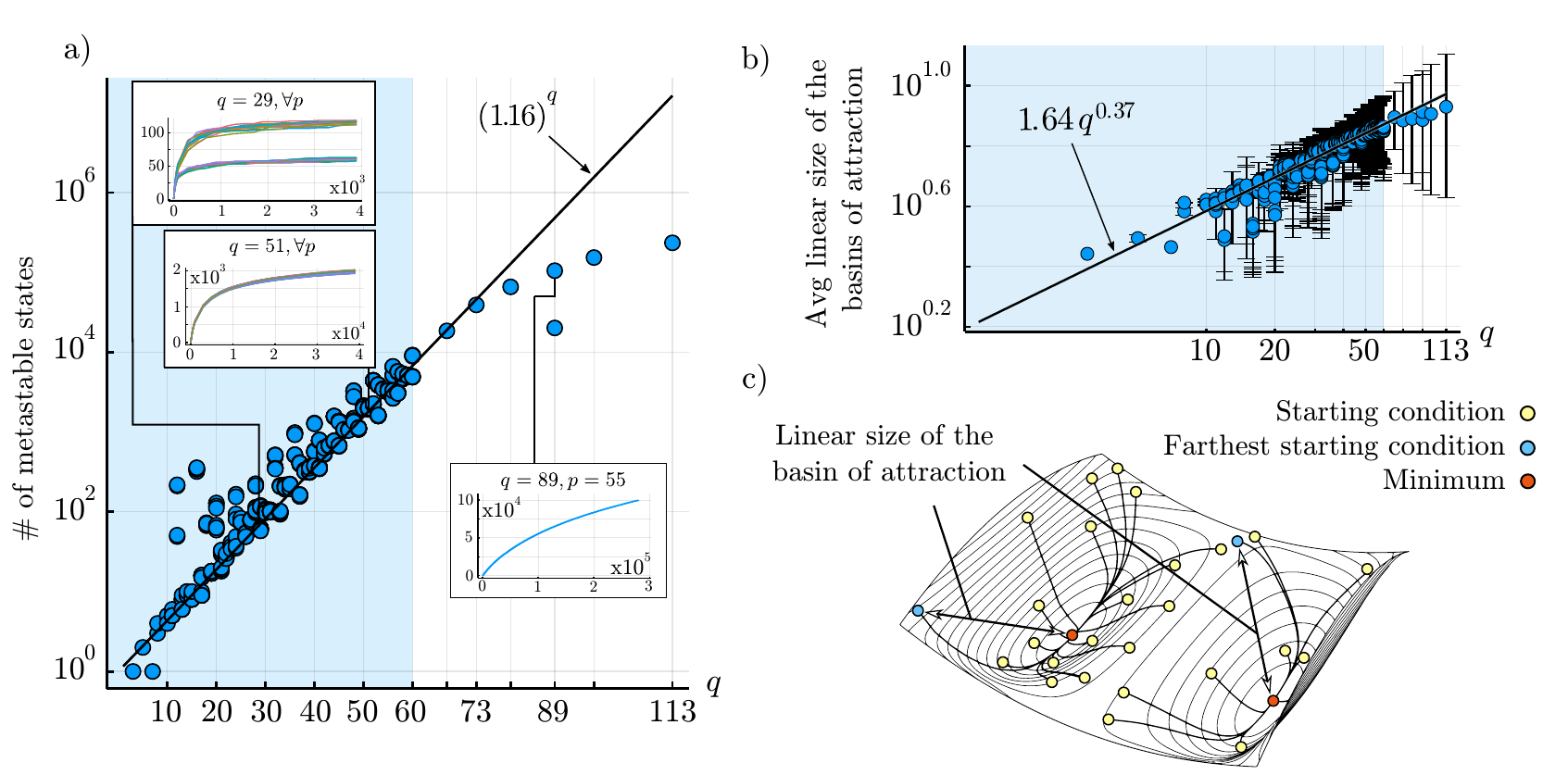}
\caption{\textbf{Exponential number of metastable states increasing $q$.}
(a) Scaling of the number of local minima of the energy in Eq. \eqref{eq. energy} as a function of $q$ (different values of $p$ are shown for each value of $q$; see SM). To check whether at a given $q$ the estimate of the number of metastable states is reliable, in the insets we display the number of distinct minima found as a function of the number of gradient descents attempted for three representative values of $q$. The closer we are to saturation, the more accurate is the estimate. Above $q \sim 60$ the estimate is not reliable, since we are missing a large number of minima (see right inset). The log-linear scale of the main plot highlights the exponential growth below $q \sim 60$, where the estimate is reliable (see left insets).
(b) The average linear size of the basins of attraction at fixed $\alpha=p/q$ grows algebraically with $q$ as $q^{0.37}$. Error bars denote one sample variance.
(c) The linear size of a given basin of attraction is measured by collecting all the initial conditions that flow to a given minimum and keeping as the linear size the distance between the minimum and the farthest initial condition from it. 
}
\label{fig:2}
\end{figure*}

In conclusion, the variational density of free-energy in the thermodynamic limit is given by: 
\begin{equation}
\label{eq. free_energy}
f(\{\tilde m_r\})=1/(2\ell q) [E (\{\tilde m_r\}) + 1/\beta S(\{\tilde  m_r\})]
\end{equation} 
and the actual free-energy of the system at equilibrium is obtained by finding the global minimum $\tilde{\mathbf m}^\ast = (\tilde m_0^\ast, \tilde m_1^\ast, \dots, \tilde m_{q-1}^\ast)$ of this $q$-dimensional function. Interestingly, it turns out that one can further reduce the dimensionality of the problem from $q$ to $\lfloor q/2 +1 \rfloor$, by exploiting the the symmetry under reflection around the anti-diagonal of the $(q-1)\times (q-1)$ interaction matrix obtained by removing the first row and column of the original reduced matrix (see also Fig. \ref{fig:1} and SM \cite{SM}).
This reduction is crucial for numerical simulations as it halves the number of effective degrees of freedom of the model.
 
\emph{Exponential number of metastable states at $T=0$---} To characterize the $T=0$ behaviour of the model, we start by focusing our attention on the enumeration of the metastable states, i.e. local minima of the energy in Eq.~\eqref{eq. energy}.
As a first point, we observe a proliferation of the local minima of the energy in Eq. \eqref{eq. energy} when we increase $q$. This phenomenon is shown for two different values of $\alpha$ in Fig. \ref{fig:1}c,  where we exhibit a representation of the energy landscape based on hierarchical clustering (see Fig. \ref{fig:1} for a more detailed explanation). In the following we numerically characterize the scaling of the number of metastable states with $q$.

Again our strategy is to probe the energy landscape by performing many gradient descents starting from a very large number of different initial conditions sampled uniformly on the $q$-dimensional hypercube $|\tilde m_r| \leq 1$ $\forall r = 0,1,\cdots, q-1$, and to count the total number of different local minima identified in this way. This algorithm presents two obvious drawbacks: (i) it may fail in finding the local minima of Eq. \eqref{eq. free_energy} with very narrow basins of attraction; (ii) for large $q$ it may systematically underestimates the number of local minima if this is growing exponentially, since the number of initial conditions we need to probe the landscape also grows exponentially in $q$.
Thus, our algorithm provides a lower bound to the number of local minima at fixed $q$, and may in any case systematically miss those minima with extremely narrow basins of attraction.
As a consequence, we need to pay particular attention to understand whether the estimate of the local minima at a given $q$ we obtain is reliable or not, i.e. if the lower-bound we provide is strict or not.

An effective method to check whether at a given $q$ we have exhaustively found most of the local minima of the landscape is to monitor the number of distinct minima found as a function of the number of gradient descents performed (see insets in Fig. \ref{fig:2}a). In this way, we can immediately argue whether the estimate is reliable by looking at how close we are to saturation. It turns out that $\sim 10^5$ gradient descents are sufficient below $q \simeq 60$, whereas one should increase this number by at least one order of magnitude for $q > 60$ and this is beyond our computational power. 

Our results for the scaling of the number of local minima $N_{\mathrm{min}}$ as a function of $q$ are shown in Fig. \ref{fig:2}a. It turns out that in the regime where the estimate is reliable (up to values of $q \simeq 60$), we find an exponential proliferation of metastable states of the form $N_{\mathrm{min}} \sim b^q$ with $b \simeq 1.16$. 

We also computed the average linear size of the basins (defined as the average distance between a minimum and the furthest starting point leading to said minimum under gradient descent), which is relevant for the application to associative memories \cite{Lev2020arxiv}
. We found that these basins are extensive in the thermodynamic limit, i.e. they occupy a finite volume of the phase space, and their linear size grows with $q$ as $q^c$ with $c \simeq 0.37$. 

\begin{figure}[t]
\includegraphics{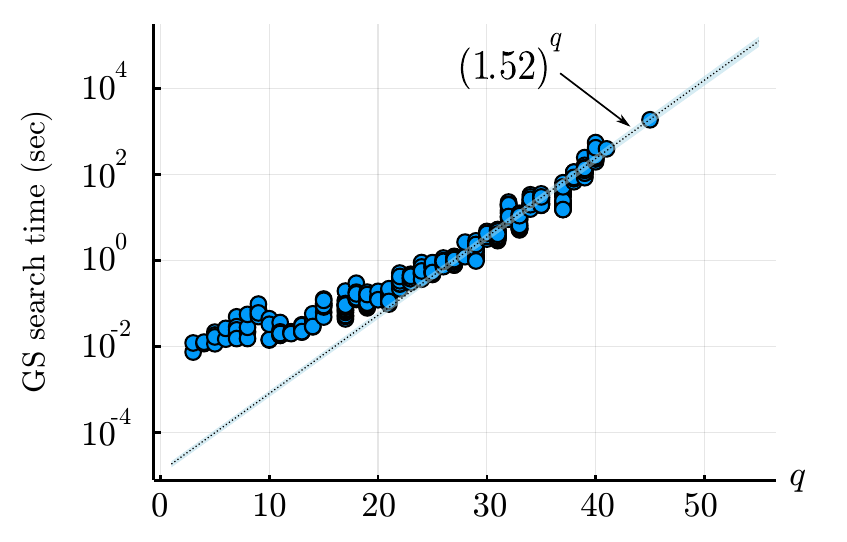}
\caption{\textbf{The run time for the ground state search problem scales exponentially with $q$.}
Scaling of the run time of the ground state (GS) search performed using a state-of-the-art optimizer (IBM CPLEX \cite{cplex2009v12, DunningHuchetteLubin2017}) as a function of $q$ (again, different values of $p$ are shown for each value of $q$). Up to $q=45$, the run time scales exponentially with $q$, suggesting that the GS search problem is a worst-case instance of non-positive-definite quadratic programming.
}
\label{fig:3}
\end{figure}

\emph{Self-induced quenched disorder in CQED---} The exponential scaling of the number of local minima, which is similar to the one observed in spin glass models \cite{MPV1986}, is due to the competition of ferromagnetic and antiferromagnetic couplings in the Hamiltonian of the model, leading to frustration. However, this result alone does not guarantee that the energy landscape of our model is complex: indeed it is well known that frustrated Hamiltonians of Ising spins may display an exponential number of degenerate ground states without exhibiting a spin glass phase, i.e. without extensive energy barriers, as in the case of the antiferromagnetic Ising model on the triangular lattice.

A significative hint suggesting that the model has a true glassy phase is given by the computational complexity of the problem of finding the model's ground state. In fact, another interesting feature of many genuine spin glass models with complex energy landscapes is that finding their ground state is an NP-hard problem (intuitively this means that the computational complexity of this optimization problem grows exponentially with the system size): this is true, for instance, for the Sherrington-Kirkpatrick (SK) and for the $p>2$-spin Ising/spherical models \cite{GARDNER1985747, Barahona_1982}, or for the $k$-SAT problem \cite{Mezard812}. Let us consider the effective energy in Eq. \eqref{eq. energy}: here we have to minimize a quadratic function of $q$ magnetizations on the domain $|\tilde m_r|$ $\forall r = 0,1,\dots, q-1$. In computer science, this is called a \emph{quadratic programming} problem \cite{998639,mezard2009information,Pardalos1991} and its computational complexity depends on whether the corresponding quadratic form is positive definite or not. In particular, the problem is NP-hard if the quadratic form has at least one negative eigenvalue. In the case of Eq. \eqref{eq. energy}, it is possible to show that the reduced interaction matrix $J_{rs} = \cos(2\pi p r s /q)$ has at least one negative eigenvalue for any $q$ (see SM \cite{SM}). 

It is worth remarking that the NP-hard nature of the combinatorial optimization problem of finding the ground state of a given model alone is not necessarily a symptom of a complex free-energy landscape. Indeed the theory of computational complexity deals with the \emph{worst case} scenario, whereas thermodynamic properties as the free-energy provide information on the \emph{typical} behavior of the system. As such, It may always be that worst case instances of an NP-hard problem have zero measure with respect to the probability distribution of the couplings/disorder.  
In our specific case, we explicitly observed an exponential scaling of the time needed to find the GS of the system using a state-of-the-art generic QP optimizer (see Fig.~\ref{fig:3}), suggesting that the problem under consideration is indeed a worst-case instance, and that the model has a low temperature glassy phase.

This last observation allows us to highlight a crucial difference between spin-spin interaction in multimodal CQED and the aforementioned systems such as the SK model: in most spin glasses, the number of couplings grows with the size of the system (e.g. in the SK model, we need to specify $N(N-1)/2$ parameters at size $N$). 
Moreover, the couplings are random variables.
On the contrary, the effective Ising model we have studied does not contain explicitly any quenched disorder, and only the parameter $\alpha$ has to be specified at any size $N$. Only a few spin models exhibiting a complex free-energy landscape without quenched disorder exist in the literature \cite{Bernasconi1987, Parisi1994JPA1, BouMez1994}. These systems have been studied extensively in the past, since they are considered toy models of the glass transition, where disorder is \emph{self-induced} and not present at the level of the Hamiltonian. It is worth noticing that interaction in CCQED provides an explicit physical realization (up to a rescaling of the coupling constants and boundary terms) of the \emph{cosine model} introduced in Ref. \cite{Parisi1994JPA2}, if one chooses $p = 1$ and $q = N$.

\emph{Discussion and Outlooks---} Recently, self-induced spin glassiness has been proposed to describe the unconventional spin glass state arising in crystalline neodymium \cite{Kamber2020Science} and predicted to occur in two- and three-dimensional arrays of Ising spins with long-range interactions \cite{Principi2016PRL, Kolmus2020}. Remarkably, the investigation of self-induced glassy phases in confocal CQED presents two major advantages over traditional solid state systems: (i) in CQED the experimental setup is highly controllable: the interaction can be tuned by manipulating the atoms positions (e.g. by using optical lattices) and the spin states can be probed by holographic imaging, as recently demonstrated in \cite{Lev2019PRL}. This is in striking contrast with standard spin glasses where the system is \emph{dirty} and quite difficult to probe. (ii) The slow timescales that are typical of a glassy dynamics should undergo a sensible speed-up in CQED due to the fast spin-flip rate, thus offering a clear way to overcome the intrinsic limitations that structural glasses usually pose. As such, our work suggests that multimodal CQED is an ideal laboratory to address the open challenges that glassy dynamics represents for modern physics.

\emph{Acknowledgements---} The authors are grateful to Giacomo Gradenigo for the useful remarks and discussions. P. R. acknowledges funding by the INFN Fellini program H2020-MSCA-COFUND Grant Agreement n. 754496.





%

\pagebreak
\clearpage

\widetext
\begin{center}
\textbf{\large Supplemental Materials}
\end{center}
\setcounter{equation}{0}
\setcounter{figure}{0}
\setcounter{table}{0}
\setcounter{page}{1}
\makeatletter
\renewcommand{\theequation}{S\arabic{equation}}
\renewcommand{\thefigure}{S\arabic{figure}}
\renewcommand{\bibnumfmt}[1]{[S#1]}

\section{Shifted chain}
The model described in the main text [Equation~\eqref{Eq.couplings}] is not translationally invariant. This requires to treat separately the case where the first atom is not exactly at the origin of the chain. Suppose that the $m-th$ atom is now in position $\mathbf{r}_m = a (m-L + s) \hat{\mathbf{n}}$, with $s\in [0,1)$ a real parameter representing a global shift $s a$ of the chain. Then the couplings are
\begin{equation}
J_{ij}(s) = \cos\left[2 \pi \alpha (i-L+s) (j-L+s) \right] \,.
\end{equation}
To find the periodicity of this matrix, we can search for a $t$ such that $J_{i+t,j}(s)=J_{i,j}(s)$. The result is that $t = u q$, with $u \in \mathbb{N}$, \emph{assuming that} also $s$ is rational, with $s = v/u$. Thus, for a fixed shift $s = v/u$, the periodicity is $uq$. The choice of scaling $L= q u \ell$ makes possible to rewrite the interaction matrix as
\begin{equation}
J_{ij}(s) = \cos\left[2 \pi \alpha (i+s) (j+s)\right]\,, 
\end{equation}
and the analysis reduces to the case without shift.

\section{Free energy at \texorpdfstring{$\alpha=\frac{a^2}{\omega_0^2}=\frac{p}{q}$}{alpha=p/q}}

In this Section we derive the expression for the free energy of the model.
The Hamiltonian of the model reads (with $N = 2L$)
\begin{equation}
    \begin{split}
        E &= \frac{1}{4L} \sum_{i,j=0}^{2L-1} \cos\left(2\pi\frac{\mathbf{r}_i\cdot \mathbf{r}_j}{w_0^2}\right) \sigma_i \sigma_j 
        = \frac{1}{4L} \sum_{i,j=0}^{2L-1} \cos\left(2 \pi \alpha (i-L) (j-L) \right) \sigma_{i} \sigma_{j}           \, ,
    \end{split}
\end{equation}
where we introduced $\alpha=a/\omega_0$.

We focus on the case of rational $\alpha$, i.e. $\alpha=p/q$ with $p$ and $q$ coprime positive integers.
Without loss of generality, we can restrict to the case $p<q$.
In fact, if $p>q$, then $p/q = n + p'/q$ for some $n \in \mathbb{Z}$, and $n$ can be dropped thanks to the periodicity of the cosine and to the fact that $i$ and $j$ are integers.

We set $L=q\ell$ with integer $\ell$ for simplicity. We will see that, when $L \neq q \ell$, the thermodynamics of the model is the same up to corrections of order $\bigO(\ell^{-1})$.

\subsection{The internal energy}

If $\alpha=p/q$, the interaction matrix $J_{i,j}$ is periodic of period $q$, i.e. $J_{i+q,j} = J_{i,j}$ and $J_{i,j+q} = J_{i,j}$. 
We can use this symmetry to greatly simplify the energy.
Let us group the spin variables into $q$ magnetizations
\begin{equation}\label{eq:E2}
    \begin{split}
        E &= \frac{1}{4q\ell} \sum_{i,j=0}^{2L-1} \cos\left(\frac{2\pi p}{q} (i-q\ell)(j-q\ell) \right) \sigma_{i} \sigma_{j}  \\
          &= \frac{1}{4q\ell} \sum_{m,n=0}^{2 \ell-1} \sum_{r,s=0}^{q-1} \cos\left(\frac{2\pi p}{q} (m q + r) (n q + s) \right) \sigma_{mq+r} \sigma_{nq+s}  \\
          &= \frac{\ell}{q} \sum_{r,s=0}^{q-1} \cos\left(\frac{2 \pi p}{q} r s \right) m_r m_s  \, ,
    \end{split}
\end{equation}
where $m_r = 1/(2\ell) \,\sum_{m=0}^{2 \ell-1} \sigma_{mq+r}$ and $r=0, \dots, q-1$.
Notice that $|m_r| \leq 1$.

The reduced interaction matrix of the magnetizations $m_r$, i.e.
\begin{equation}
    \begin{split}
        J'_{r,s} = \cos\left( \frac{2 \pi p}{q} r s \right) \, , \qquad 0 \leq r,s \leq q-1  \, ,
    \end{split}
\end{equation}
has another symmetry, namely $J'_{i,j} = J'_{q-i,j}$ and $J'_{i,j} = J'_{i,q-j}$, due to the parity of the cosine.
This allows to further simplify the energy by introducing the reduced magnetizations
$\tilde{m}_0 = m_0$, $\tilde{m}_s = (m_s + m_{q-s})/2$ for $s = 1, \dots, \lfloor (q-1)/2 \rfloor$ and, if $q$ is even, $\tilde{m}_{q/2} = m_{q/2}$.
Easy computations show that the internal energy can be rewritten as
\begin{equation}\label{eq:E3}
    \begin{split}
        E = \frac{\ell}{q} &\left[ \tilde{m}_0^2 + 2 \tilde{m}_0 \tilde{m}_{\frac{q}{2}} +4 \tilde{m}_0 \sum_{r=1}^{\lfloor (q-1)/2\rfloor} \tilde{m}_r 
        + \cos\left( \frac{\pi p q}{2} \right) \tilde{m}_{\frac{q}{2}}^2 \right. \\
                           &\left.\quad+ 4 \tilde{m}_{\frac{q}{2}} \sum_{r=1}^{\lfloor (q-1)/2 \rfloor} \cos\left( \pi p r \right) \tilde{m}_r 
                               + 4 \sum_{r=1}^{\lfloor (q-1)/2 \rfloor}\sum_{s=1}^{\lfloor (q-1)/2 \rfloor}\cos\left( \frac{2 \pi p}{q} r s  \right) \tilde{m}_r \tilde{m}_s 
                           \right] 
                           \, ,
    \end{split}
\end{equation}
where $m_{q/2} = 0$ if $q$ is odd. 
Thus, the internal energy is again a quadratic form in the new magnetizations with interaction matrix $J''_{r,s}$ of the form
\begin{equation}
    \begin{split}
        J'' = 
        \begin{bmatrix}
            1      & 2         & \dots  & 2             & 1             \\
            2      & J'_{1,1}  & \dots  & J'_{1,q'}     & 2 (-1)^p      \\
            \vdots & \vdots    & \ddots & \vdots        & \vdots        \\
            2      & J'_{q',1} & \dots  & J'_{q',q'}    & 2 (-1)^{p q'} \\
            1      & 2 (-1)^p  & \dots  & 2 (-1)^{p q'} & \cos\left( \pi p q /2 \right)               \\
        \end{bmatrix}
        \, 
    \end{split}
\end{equation}
where $q' = \lfloor (q-1)/2 \rfloor$, and the last row/column are there only if $q$ is even.

In the main text we present Equation~\eqref{eq:E2} since it is more compact and does not depend on the parity of $q$.
In the numerical simulations we used Equation~\eqref{eq:E3} as it has roughly half the degrees of freedom with respect to Equation~\eqref{eq:E2}.

We conclude this section by pointing out that, for $L\neq q\ell$, the only change in our computations is that some of the $\{\tilde{m}_s\}$ are the sum of $2\ell+1$ degrees of freedom instead of just $2 \ell$.
Thus, all equations hold at leading order in a large $\ell$ expansion, with corrections of order $\bigO(\ell^{-1})$.

\subsection{The entropy}

The entropy of our model depends on which representations we would like to use, either Equation~\ref{eq:E2} or Equation~\ref{eq:E3}.
In both cases, the total entropy $S$ is given by the sum of the individual entropies of the magnetizations $\{m_s\}$ or $\{\tilde{m}_s\}$.

The individual entropy of a magnetization $m$ composed by $n$ microscopic degrees of freedom (spins) is given by
\begin{equation}
    \begin{split}
        S_n(m) = \binom{n}{\frac{n}{2}(m+1)} = n \log 2 - \frac{n}{2} \left[ (1+m)\log(1+m) + (1-m)\log(1-m) \right] + \bigO(\log(n)) \, 
    \end{split}
\end{equation}
that is the number of configurations with $n (m+1)/2$ spins up, i.e. in which the magnatization equals $m$. 
The big-O notation refers to the thermodynamic limit $n\rightarrow\infty$.

Thus, the entropy associated with the internal energy given in Equation~\eqref{eq:E2} equals (in the large $\ell$ limit)
\begin{equation}
    \begin{split}
        S(\{\tilde{m}_s\}) 
        = \sum_{s=0}^{q-1} S_{2\ell}(\tilde{m}_s) 
        = 2\ell \left[ q \log 2 - \sum_{s=0}^{q-1} \left( \frac{1+\tilde{m}_s}{2}\log(1+\tilde{m}_s) +\frac{1-\tilde{m}_s}{2}\log(1-\tilde{m}_s) \right) + \bigO\left( \frac{\log(\ell)}{\ell} \right)  \right]
        \, .
    \end{split}
\end{equation}

On the other hand, the entropy associated with the internal energy given in Equation~\eqref{eq:E3} equals (in the large $\ell$ limit)
\begin{equation}
    \begin{split}
        S(\{m_s\}) 
        &= S_{2\ell}(m_0) + S_{2\ell}(m_{\frac{q}{2}}) + \sum_{s=1}^{\lfloor\frac{q-1}{2}\rfloor} S_{4\ell}(m_s) \\
        &= 2\ell \left[ q \log 2 
            - \frac{1-m_0}{2}\log(1-m_0) - \frac{1+m_0}{2}\log(1+m_0) \right.\\
        &\quad- \frac{1-m_{\frac{q}{2}}}{2}\log(1-m_{\frac{q}{2}}) - \frac{1+m_{\frac{q}{2}}}{2}\log(1+m_{\frac{q}{2}}) 
       \\
        &\quad\left.
        - \sum_{s=1}^{\lfloor\frac{q-1}{2}\rfloor} \Big( (1+m_s)\log(1+m_s)+(1-m_s)\log(1-m_s) \Big) + \bigO\left( \frac{\log(\ell)}{\ell} \right)  \right]
        \, ,
    \end{split}
\end{equation}
where the terms depending on $m_{\frac{q}{2}}$ must be discarded if $q$ is odd.

Again, in the case $L\neq q\ell$, the corrections to the entropies are of order $\bigO(\ell^{-1})$.

\section{The spectrum of the reduced interaction matrix}

The reduced interaction matrix is defined as
\begin{equation}
    \begin{split}
        J'_{r,s} = \cos\left( \frac{2 \pi p}{q} r s \right)  \, , \qquad 0 \leq r,s, \leq q-1 \, .
    \end{split}
\end{equation}
As we have already seen, this matrix has additional symmetries left. 
Nonetheless, it is more tractable than the fully reduced one, i.e. the interaction matrix $J''$ for the magnetizations $\{\tilde{m}_s\}$ in Equation~\ref{eq:E3}.

Here we characterize the spectrum of $J'$.
First of all, we notice that $(J')^2$ is given by:
\begin{equation}
    \begin{split}
        (J')^2_{i,j} 
        &= \sum_{k=0}^{q-1} J'_{i,k} J'_{k,j}
        = \sum_{k=0}^{q-1} \cos\left( \frac{2\pi p}{q} i k \right) \cos\left( \frac{2\pi p}{q} k j \right) 
        = \frac{1}{2} \sum_{k=0}^{q-1} \left[ \cos\left( \frac{2\pi p}{q} (i+j) k \right) + \cos\left( \frac{2\pi p}{q} (i-j) k \right) \right]
        \, .
    \end{split}
\end{equation}
Using the following identity
\begin{equation}
    \begin{split}
        \sum_{s=0}^{q-1} \cos\left( \frac{2\pi p}{q} r s \right) 
        = \Re \left[ \sum_{s=0}^{q-1} e^{i \frac{2\pi p r}{q} s} \right]  
        = \Re \left[ \frac{1-e^{i 2 \pi p r}}{1-e^{i \frac{2 \pi p r}{q}}} \right]  
        = q \delta_{r,0}
        \, ,
    \end{split}
\end{equation}
we obtain
\begin{equation}
    \begin{split}
        (J')^2_{i,j} = \frac{q}{2} \left( \delta_{i,j} + \delta_{i,-j} \right) 
         \, .
    \end{split}
\end{equation}
As such, the spectrum of $(J')^2$ is given by:
\begin{itemize}
    \item the eigenvector $(1,0,\dots,0)^T$, with eigenvalue $q$;
    \item the $\lceil (q-1)/2 \rceil$ eigenvectors $(0,1,0,0,\dots,0,0,1)^T$, $(0,0,1,0,\dots,0,1,0)^T$ and so on, with eigenvalue $q$;
    \item the $\lfloor (q-1)/2 \rfloor$ eigenvectors $(0,1,0,0,\dots,0,0,-1)^T$, $(0,0,1,0,\dots,0,-1,0)^T$ and so on, with eigenvalue $0$.
\end{itemize}
Thus, $J'$ has $\lfloor (q-1)/2 \rfloor$ null eigenvalues, and $\lceil (q-1)/2 \rceil +1$ eigenvalues $\pm \sqrt{q}$.
We notice that the null eigevalues correspond to the additional symmetries of $J'$, and will not be part of the spectrum of $J''$.

Moreover, $J'$ has always at least one negative eigenvalue. 
In fact, it is straightforward to check that the vector $v = (1-\sqrt{q},1, \dots, 1)^T$ is an eigenvector of $J'$ with eigenvalue $-\sqrt{q}$:
\begin{equation}
    \begin{split}
        \left(J'v\right)_r 
        = (1-\sqrt{q}) + \sum_{s=1}^{q-1} \cos\left( \frac{2\pi p}{q} r s \right) 
        = -\sqrt{q} + \sum_{s=0}^{q-1} \cos\left( \frac{2\pi p}{q} r s \right) 
        = -\sqrt{q} + q \delta_{r,0}
        = - \sqrt{q} v_r
         \, .
    \end{split}
\end{equation}
Notice that due to the symmetry in the components $v_r$ with $r = 1, \dots, q-1$ this remark translates to $J''$ as well. 

\section{High-temperature phase}

In the large temperature limit $T\rightarrow\infty$, the free energy landscape is dominated by the entropic contribution. The free energy is factorized in each magnetization, and its minimum can be computed by minimizing each of the entropies. This gives a paramagnetic high-temperature phase characterized by the conditions $m\equiv0$  or, equivalently, $\tilde{m}\equiv0$.
Around the paramagnetic minimum, the free energy at finite temperature $T$ can be expanded in Taylor series as
\begin{equation}
    \begin{split}
        \frac{1}{\ell}(E - T S )
        = \frac{1}{q} \sum_{r,s=0}^{q-1} J'_{r,s} m_r m_s 
        - 2 q T \log2  
        + T \sum_{s=0}^{q-1} m_s^2 + \bigO\left(m^3,\log(\ell)\right)
        \, .
    \end{split}
\end{equation}
Thus, the stability of the paramagnetic minimum is determined by the sign of the eigenvalues of the matrix $\frac{1}{q} J' + T \mathbb{1}_q$, where $\mathbb{1}_q$ is the $q\times q$ identity matrix.
The minimum eigenvalue is given by $-\frac{1}{\sqrt{q}} + T$, giving a critical temperature for the paramagnetic stability $T_{\rm param} = \beta_{\rm param}^{-1} = \frac{1}{\sqrt{q}}$.

Preliminary numerical investigation (to be reported elsewhere) suggest that this is a second-order phase transition.

\section{Details of numerical simulations}

We performed extensive numerical simulations of the model to characterize its zero-temperature energy landscape.
All simulations were performed using the fully reduced model given in Equation~\ref{eq:E3}.

The code to reproduce our results is available on Github at the link \url{https://github.com/vittorioerba/FreeEnergyMultimodeQEDpaper}.

\subsection{Metastable states}

To sample the minima of the energy landscape described by Equation~\ref{eq:E3} we used a simple steepest descent algorithm with update rule
\begin{equation}
    \begin{split}
        m(t+1) = m(t) - \eta \nabla E\left( m(t) \right)  \, 
    \end{split}
\end{equation}
and step-size $\eta = 10^{-3}$.
We implemented manually the box constraint by:
\begin{itemize}
    \item setting to 1 or to $-1$ each magnetization such that $m_i(t) \in [-1,1]$ and $m_{i+1}(t) \notin [-1,1]$;
    \item setting to 0 all the components $i$ of the gradient that point out from the hypercube when $m_i = \pm 1$.
\end{itemize}
Finally, we set the stopping condition to
\begin{equation}
    \begin{split}
        || \nabla E\left( m(t) \right) ||_2 < \tau  \, 
    \end{split}
\end{equation}
with threshold $\tau = 10^{-3}$.
All optimization runs that that failed to reach this condition before $t_{\rm max} = 10^6$ steps where discarded, as well as all other simulations that failed for any other reason.

For each value of $p$ and $q$, we performed $N_{\rm run}(p,q)$ gradient descents mainly based on the value of $q$:
\begin{itemize}
\item for $11 \leq q \leq 20$, we performed $N_{\rm run} \geq 1 \times 10^3$;
\item for $21 \leq q \leq 29$, we performed $N_{\rm run} \geq 4 \times 10^3$;
\item for $30 \leq q \leq 40$, we performed $N_{\rm run} \geq 15 \times 10^3$;
\item for $41 \leq q \leq 50$, we performed $N_{\rm run} \geq 25 \times 10^3$;
\item for $41 \leq q \leq 60$, we performed $N_{\rm run} \geq 40 \times 10^3$.    
\end{itemize}
For these values of $q$ we studied all the values of $p<q$ comprime with $q$.
Finally, for $q \geq 61$, we performed $N_{\rm run} \geq 150 \times 10^3$ gradient descents for a very restricted set of values of $p$.
The precise number $N_{\rm run}(p,q)$ simulated for each pair $(p,q)$ is reported in the repository hosting the code.

To assess whether for a given value of $p/q$ all local minima had been enumerated, we studied the number of distinct local minima found $N_{p,q}(n)$ as a function of the number of gradient descents performed $n$.
As the model has an obvious $\mathbb{Z}_2$ symmetry left, we counted each minimum $m^*$ and his opposite $-m^*$ as the same minimum.
When $n$ is small, under the hypothesis that the basins of attraction of the local minima are somewhat comparable in size, we expect that $N_{p,q}(n) \sim n$, as each new descent finds an unknown minimum.
When $n$ is large, we expect that $N_{p,q}(n)$ saturates to a constant value $N_{p,q}(\infty)$, as all local minima were already visited by previous descents.

\begin{figure}[t]
\includegraphics{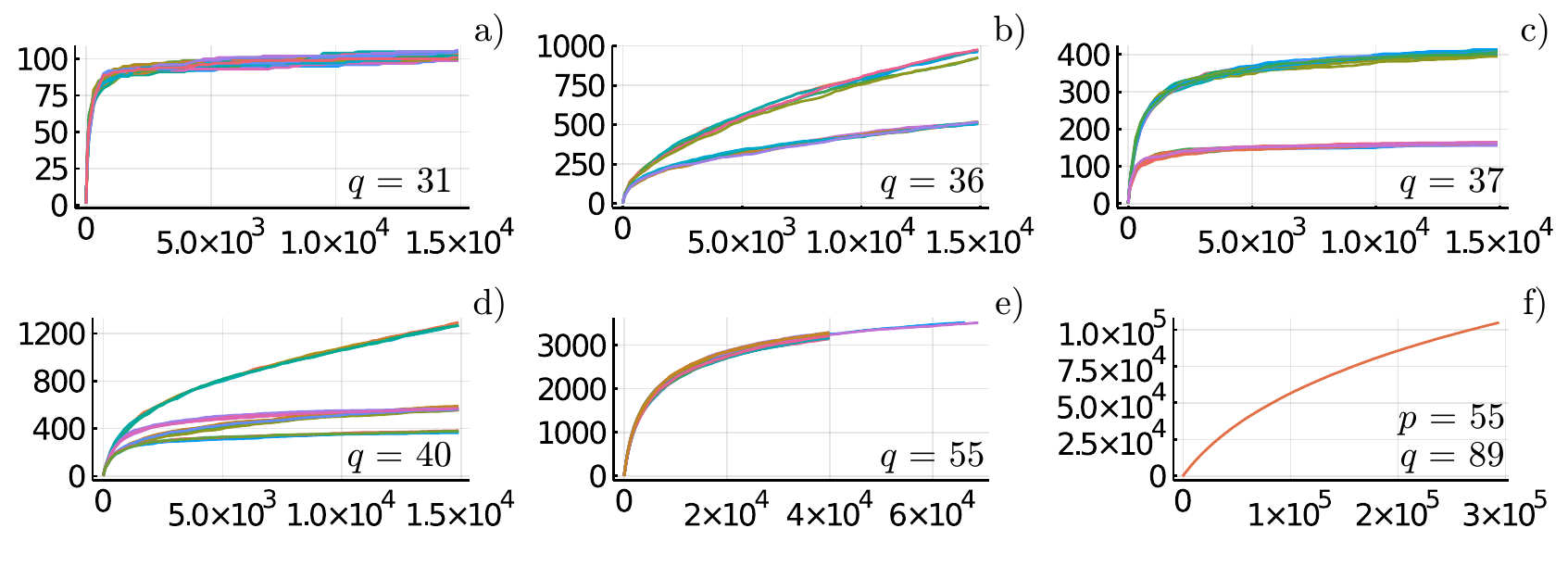}
\caption{\textbf{Behaviour of $N_{p,q}(n)$ for a selection of values of $p$ and $q$.}
} 
\label{fig:SM1}
\end{figure}

Figure~\ref{fig:SM1} shows the behaviour of $N_{p,q}(n)$ extracted from the simulations for a selection of values of $q$.
Firstly we observe that, at fixed $q$, the curves $N_{p,q}(n)$ may or may not collapse onto a single curve.
In the latter case, one may expect that each value of $p$ has its own $N_{p,q}(n)$ curve. 
Suprisingly, there exists only a limited number of possible curves over which the experimental $N_{p,q}(n)$ distribute.
Another important observation is that there are values of $q$, for example $q=36$, where the curves $N_{p,q}(n)$ are still far from saturation, while for larger values of $q$, for example $q=37$, saturation is reached in the same number of descent runs $n$.
We notice that for those values of $q$ where the curves $N_{p,q}$ do not collapse onto a single curve, see for example $q=40$, we obtain that for particular values of $p$, $N_{p,q}(n)$ saturates, while for others, $N_{p,q}(n)$ do not saturate in the same number of descent runs $n$.

Finally, we notice that, even if $N_{p,q}(n_{\rm run})$ is not a good descriptor  of $N_{p,q}(\infty)$, it does provide a lower bound for it. 

\subsection{Ground states}

To compute the ground states of the model we used the CPLEX \cite{cplex2009v12} optimization library, which provides a solver for non-positive-definite quadratic programming problems. 
To interface with the solver, we used the Julia API \cite{DunningHuchetteLubin2017}.
See the Github repo for the details of the implementation.

\end{document}